# Two polymorphs of a new intermetallic $Ce_2Rh_2Ga$ – crystal structure and physical properties

Sergey Nesterenko[1], Anna Tursina[1], Mathieu Pasturel[2], Sindisiwe Xhakaza[3], André Strydom[3,4]

[1] *Department of Chemistry, Lomonosov Moscow State University, 119991 Moscow, Russia*

[2] *Univ Rennes, CNRS, ISCR – UMR6226, F-35000 Rennes, France*

[3] *Highly Correlated Matter Research Group, Physics Department, University of Johannesburg, PO Box 524, Auckland Park 2006, South Africa*

[4] *Max Planck Institute for Chemical Physics of Solids, 40 Nöthnitzerstr., D-01187 Dresden, Germany*

**Highlights**

• $Ce_2Rh_2Ga$ crystallizes in $Pr_2Co_2Al$-type (LT) and $La_2Ni_3$-type (HT) structures.

• The transition occurs at 864(5)°C

• Both structures can be described by different stacking of Ce-centered $CeGa_4$ building blocks.

• HT-$Ce_2Rh_2Ga$ exhibits a phase transition, putatively of magnetic nature, at the unusually high temperature of 128.5 K

**Abstract**

We report on the synthesis and physical properties of the new ternary intermetallic $Ce_2Rh_2Ga$. Its formation and dimorphism have been investigated by powder and single crystal X-ray diffractometry, as well as by differential thermal analyses. LT- and HT-$Ce_2Rh_2Ga$ can be obtained by the chosen respective thermal treatment. The high-temperature form (HT-$Ce_2Rh_2Ga$), stable above 864(5)°C, is orthorhombic $La_2Ni_3$-type [*Cmce*, *oS*20, $a$=5.851(2) Å, $b$=9.618 (2) Å, $c$=7.487 (2) Å, $V$=421.3(2) Å$^3$]; the low-temperature form (LT-$Ce_2Rh_2Ga$), stable below this temperature, is monoclinic $Pr_2Co_2Al$- ($Ca_2Ir_2Si$-) type [*C*2/*c*, *mS*20, $a$=10.0903(6) Å, $b$=5.6041(3) Å, $c$=7.8153(4) Å, $\beta$=104.995(3)°, $V$=426.88(4) Å$^3$]. The magnetic measurements were conducted on two different samples, namely on the LT-$Ce_2Rh_2Ga$ and the other on the HT-$Ce_2Rh_2Ga$ sample. The HT-$Ce_2Rh_2Ga$ compound is found to exhibit a phase transition at 128.5 K. By virtue of the sharp peak anomaly in the temperature dependence of the magnetic susceptibility this phase transition is plausibly of antiferromagnetic origin, but occurs at a remarkably high temperature compared to what is typical for Ce binary and ternary compounds. LT- $Ce_2Rh_2Ga$ has a ground state of ferromagnetic nature which sets in at a paramagnetic-to-ferromagnetic phase transition at 5.2 K.

**1. Introduction**

The $R_2T_2X$ series of compounds (with R = rare earth, T = transition metal and X = $p$-metal), the so-called 2:2:1 family, are known to exhibit a wide variety of physical properties at low temperature. Their crystallographic characteristics and magnetic data have been summarized in review articles (see *e.g.* [1-3]). In particular, in the ternary systems with X = In, the known $R_2T_2X$ phases show ground states ranging from magnetically ordered Kondo compounds for T = Pd, Cu, and Ag via a heavy fermion behavior in $Ce_2Pt_2In$ to intermediate valence for T = Ni and Rh [4]. The $Ce_2Ni_2Sn$ stannide behaves like a Kondo system and orders antiferromagnetic below $T_N$ = 4.7 K [5]. The cerium-based intermetallics $Ce_2Ni_{1.88}Cd$ [6, 7] as well as $Ce_2Ni_2Ga$ [8] exhibit a fluctuating valence. $Yb_2Ni_2Al$ and $Yb_2Pd_2Sn$ were reported as heavy fermion systems with magnetically non-ordered ground states [9, 10].

Most of $R_2T_2X$ phases crystallize in the four structure types: $Mo_2B_2Fe$ (tetragonal, $P4/mbm$) [11], $Er_2Au_2Sn$ or $U_2Pt_2Sn$ (tetragonal, $P4_2/mnm$) - a superstructure of $Mo_2B_2Fe$ [12, 13], $W_2B_2Co$ (orthorhombic, $Immm$) [14], and $Mn_2B_2Al$ (orthorhombic, $Cmmm$) [15].

Five less common types are encountered in literature. Two of these are monoclinic ones: HT-$Pr_2Co_2Al$ or $Ca_2Ir_2Si$ ($C2/c$) - stacking variant of the $W_2B_2Co$ [16, 17] - and LT-$Nd_2Cu_2Cd$ ($C2/c$, own structure type) - superstructure of $Mn_2B_2Al$ [18]; the other three structures are orthorhombic: $La_2Ni_3$ ($Cmce$) [19], $o$-$La_2Ni_2In$ ($Pbam$) [20], and $Ca_2Pd_2Ge$ ($Fdd2$) [21]. Only a few representatives of these types are reported to date [22-26].

The structural diversity of the $R_2T_2X$ family is frequently accompanied by polymorphic phenomena between these structure types. A structural transformation from $Mo_2B_2Fe$ to $W_2B_2Co$-type was observed in the $R_2Ni_2Sn$ (R = Ce, Pr, Nd, Sm, Gd - Tm, Lu) system as a function of the rare-earth size or external pressure [27, 28]. Also, depending on the size of the rare earth element, several groups of indides and stannides crystallize with the $Mo_2B_2Fe$-type or the $Er_2Au_2Sn$ structure [29, 30]. A transition from the $Mo_2B_2Fe$-type to the $Mn_2B_2Al$-type, provided by geometrical reasons or temperature, was found in the $RE_2Ni_2X$ (X = In or Cd) series. The latter transition is of a reconstructive nature [31, 32].

Polycrystalline $Mo_2FeB_2$-type $U_2Pt_2In$ transforms via a displacive phase transition to the $Er_2Au_2Sn$-type in the process of single crystal growth [33]. $Pr_2Co_2Al$ is dimorphic with a $W_2B_2Co$-type low-temperature modification and a monoclinic (space-group $C2/c$) high-temperature modification [16]. According to [20], $La_2Ni_2In$ crystallizes with the orthorhombic (*Pbam*) or tetragonal Mo$_2$B$_2$Fe (*P4/mbm*) modification depending on annealing temperature. Interestingly, the same transformation takes place in the course of the hydrogenation of La$_2$Ni$_2$In, during which the symmetry changes from tetragonal to orthorhombic [34]. Finally, in the Gd-Co-Ga system, the compound with the W$_2$B$_2$Co-type structure is formed at the stoichiometric composition, but the Ga-deficient compound with Gd$_2$Co$_{2.9}$Ga$_{0.1}$ composition crystallizes in the Mn$_2$B$_2$Al-type structure [35, 36], and in the Dy-Ni-Si system the variation of alloy composition from Dy$_2$Ni$_{2.35}$Si$_{0.65}$ to Dy$_2$Ni$_{2.5}$Si$_{0.5}$ leads to crystallization of W$_2$B$_2$Co- or La$_2$Ni$_3$-type compounds [36].

In the course of our investigations on the ternary Ce-Rh-Ga system, we have discovered a new compound, namely Ce$_2$Rh$_2$Ga. As it turned out, it is dimorphic, with the high temperature form adopting the orthorhombic La$_2$Ni$_3$-type structure and the low temperature form crystallizing in the monoclinic Pr$_2$Co$_2$Al-type, and shows a new type of phase transition in the R$_2$T$_2$X family. Synthesis, crystal structures determination and LT physical properties for Ce$_2$Rh$_2$Ga polymorphs and La$_2$Rh$_2$Ga are reported herein. Preliminary data were presented at the SCTE 2018 conference [37].

**2. Experimental**

The alloys were prepared by arc-melting (Arc Melter AM/0.5, Edmund Bühler GmbH) appropriate amounts ((Ce or La)40Rh40Ga20 (at.%)) of the constituent elements under a high-purity argon atmosphere in a water-cooled copper hearth. The purities of the starting materials were 99.85 % for Ce and La, 99.95 % for Rh, and 99.999 % for Ga. The ingots were remelted three times to ensure homogeneity. Then both the resulting cerium and lanthanum ingots were divided into two parts. First halves of Ce based and La based samples were annealed at 700 and the second ones at 900°C for 30 days in evacuated quartz tubes and then quenched in cold water.

Both the composition and the homogeneity of the annealed samples were checked by microprobe analysis using a Carl Zeiss LEO EVO 50XVP scanning electron microscope equipped with an INCA Energy 450 (Oxford Instruments) EDX-spectrometer.

X-ray powder diffraction was performed at room temperature with the use of a STOE STADI-P transmission diffractometer with CuK$\alpha_1$ radiation ($\lambda$ = 1.54056 Å) in the 10 - 90° 2θ range with a step size of 0.01°.

Single crystals of HT-Ce$_2$Rh$_2$Ga and La$_2$Rh$_2$Ga were separated from the respective as-cast samples, whereas single crystal of LT-Ce$_2$Rh$_2$Ga was found in the annealed at 700°C sample. Single crystal X-ray diffraction data were collected at room temperature either on a Bruker Apex II (Mo *Kα* radiation) or on a CAD4 Enraf Nonius (Ag *K$_\alpha$* radiation) diffractometers. The crystal structures were refined by the full-matrix least-squares method in the anisotropic approximation for all atoms using either the SHELXL-97 software [38] or SHELXL-2015/1 one. Further details of the crystal structure investigations are listed in Table 1 and may be obtained from the Fachinformationszentrum Karlsruhe, 76344 Eggenstein-Leopoldshafen, Germany (Fax: +49-7247-808-666; E-Mail: crysdata@fiz-karlsruhe.de, http://www.fiz-karlsruhe.de/request_for_deposited_data.html) on quoting the depository numbers CSD 434348 (LT-$Ce_2Rh_2Ga$) and CSD 434386 (HT-$Ce_2Rh_2Ga$). The crystal structure of $La_2Rh_2Ga$ was deposited to the joint CCDC/FIZ Karlsruhe database via www.ccdc.cam.ac.uk with the reference number 1941455. Final atomic coordinates and equivalent isotropic displacement parameters are listed in Table 2, and the most important bond distances for $Ce_2Rh_2Ga$ polymorphs are given in Table 3.

The annealed samples of $Ce_2Rh_2Ga$ were analyzed by differential thermal analysis (DTA) using a NETZSCH Leading Thermal Analysis STA 449 F1 Jupiter Platinum RT apparatus. The polycrystalline samples (~30 mg) were put in a small $Al_2O_3$ crucible, and the subsequent measurement was conducted under a continuous He flow of 30 mL min$^{-1}$. The cooling and heating rates were 20 °C min$^{-1}$. An empty $Al_2O_3$ crucible was used as a reference.

The magnetic susceptibility and heat capacity were measured as function of temperature in high resolution using a Dynacool Physical Properties Measurement System from *Quantum Design* (San Diego, USA). In the case of magnetic susceptibility, data were recorded in both warming and cooling modes in order to investigate the reversibility of the phase transition.

## 3. Results
### 3.1 Phase analysis

Microprobe analysis on both annealed $Ce_2Rh_2Ga$ samples confirmed the chemical composition and homogeneous distribution of the elements with negligible amount of CeRhGa phase [39]. As this equiatomic ternary phase shows intermediate valent behavior without magnetic ordering [40, 41] we expect its presence in our magnetic study on $Ce_2Rh_2Ga$ to be of no consequence.

The X-ray powder patterns of these samples were different depending on the thermal treatment used, and at the same time the crystal structure of the sample annealed at 900°C does not differ from the as-cast ones. DTA was used to establish the reason for this difference (Fig. 1, and Fig. S1 in Supplementary). When heating of the sample annealed at 700°C, one relatively weak endothermic peak was detected at 864(5)°C, that may be attributed to the low temperature (LT) – high temperature (HT) polymorphous transformation, followed by melting at 970(5)°C. It is worth noticing that the reversible polymorphic reaction is not detected in the cooling branch of DTA scan; this could be due to the small amount of heat and slow kinetics associated to this reaction.

The XRD data refinement of polycrystalline samples reveals that the sample annealed at 700°C consisted solely of the LT phase, whereas the as-cast samples as well as the one annealed at 900°C contain exclusively the HT phase (Fig. 2). Other phases were not detected. The Rietveld refinements were performed with the MRIA program [42] using the crystallographic data obtained from the single crystal XRD experiments (Tables 1, 2) for orthorhombic HT-$Ce_2Rh_2Ga$ and for monoclinic LT-$Ce_2Rh_2Ga$.

**3.2 Crystal structures**

The HT-$Ce_2Rh_2Ga$ crystallizes with the orthorhombic $La_2Ni_3$-type structure [19], while the low-temperature polymorph adopts the monoclinic $Pr_2Co_2Al$ (also known as $Ca_2Ir_2Si$) type [16, 17]. The non-f-electron counterpart $La_2Rh_2Ga$ was found to crystallize with the orthorhombic $La_2Ni_3$-type structure.

The coordination polyhedra of all the atoms are presented in Fig. 3 and Fig. S2 in the Supplementary Materials. The environment of Ce atoms in the two polymorphs is rather similar and comprises six Rh, four Ga and five Ce atoms. The resulting polyhedron can be regarded as a derivative of a pentagonal prism centered on one basal face and on all side faces. In both structures, the Rh atom is coordinated by six cerium atoms and two gallium atoms in the form of a distorted square antiprism with two additional rhodium atoms capping the basal faces of the prism. Finally, the arrangement of the eight cerium atoms and four rhodium atoms around Ga results in a severely distorted icosahedron.

The Ce-Ce distances are ranging from 3.43 to 3.70 Å for the HT polymorph - within the similar range observed in the isostructural $Ce_2Ru_2Al$ with $d_{Ce-Ce}$ =3.32 and 3.64 Å [25] - and from 3.61 to 3.71 Å for the LT one (Table 3). These distances correspond well to twice the metallic radius of this element ($r_{Ce}$ = 1.825 Å [43]). Likewise, the interatomic distances Ce-Rh and Ce-Ga in HT modification are in the interval of corresponding values in the LT modification. Nevertheless, consideration of the Ga and Rh metallic radii ($r_{Ga}$ = 1.411 Å and $r_{Rh}$ = 1.345 Å [43]) highlights short Ce-Rh distances and rather elongated Ce-Ga ones. The Rh-Ga contacts of 2.56 Å in HT-$Ce_2Rh_2Ga$ and from 2.58 to 2.61 Å in the LT-$Ce_2Rh_2Ga$ are significantly smaller than the sum of the metallic radii, also suggesting a strong bonding between these elements. Opposite, more or less elongated Rh-Rh distances are observed and Ga atoms are well separated from each other and produce non-bonding contacts in both phases.

In the new compounds, the four Ga neighbors around each Ce atom form a slightly distorted $CeGa_4$ tetrahedron with Ce-Ga contacts being in the range of 3.34 – 3.39 Å and 3.30 – 3.35 Å for HT- and LT-$Ce_2Rh_2Ga$, respectively. Cerium centered gallium tetrahedra can be regarded as the main common building blocks. In the HT-Ce$_2$Rh$_2$Ga, CeGa$_4$ tetrahedra are merged by edge-sharing with two neighboring ones forming rows alternating in a ''checkerboard'' pattern running along the *a*-axis (Fig. 4a). Each CeGa$_4$ tetrahedron has three more shared edges with adjacent rows.

The CeGa$_4$ tetrahedra are delimiting tetrahedral voids centered by Rh atoms (Fig. 4b). Thus the Ga environment of Rh atoms is a RhGa$_4$ tetrahedron which is severely distorted in contrast to that of Ce-centered. Within the Rh-centered tetrahedron, two Rh-Ga contacts are bonding – each of 2.555 Å - and two others are approximately 2 Å longer – 4.540 Å and 4.862 Å. Each RhGa$_4$ tetrahedron contacts with six neighbors by edge-sharing generating a 3D framework. The superposition of the Ce-centered tetrahedra and the Rh-centered tetrahedra nets results in a three-dimensional network by sharing triangular faces composed of Ga atoms (see Fig. 4c).

In the LT-Ce$_2$Rh$_2$Ga, the CeGa$_4$-tetrahedra framework is built by edge-sharing of each of the six edges of a CeGa$_4$ tetrahedron (Fig. 5a) which is delimiting octahedral voids centered by a Rh-Rh dumbbell ($d_{Rh-Rh}$ = 2.75 Å). In this structure, each Rh atom resides in a distorted square-based pyramid RhGa$_5$ with Rh-Ga interatomic distances of 2.578 Å, 2.607 Å, 4.579 Å, 4.633 Å, and 4.668 Å. Two neighboring RhGa$_5$ pyramids share their basal faces to form a Rh$_2$Ga$_6$ octahedron (Fig. 5b). The resulting structure is produced by the CeGa$_4$- and Rh$_2$Ga$_6$-networks linked by the mutual triangular faces of the Ce-centered tetrahedra and of the Rh-double-centered octahedra (Fig. 5c).

### 3.3 Physical properties

*LT-Ce$_2$Rh$_2$Ga*. Figure 6 shows the variation of the Ce-molar magnetic susceptibility of LT-Ce$_2$Rh$_2$Ga with temperature on a logarithmic temperature scale. Below 400 K $\chi(T)$ assumes small values but follows a regular Curie-Weiss law, $\chi(T)=C/(T-\theta_P)$ down to 100 K as is shown by the line fitted to the experimental data (inset to Fig. 6). Here $C =N\mu_{eff}^2/Mk_B$ in terms of Avogadro's number $N$, the substance molar mass $M$ and Boltzmann's constant $k_B$. The fit parameters are the Weiss temperature $\theta_P$ and the effective magnetic moment $\mu_{eff}^2=g_J^2J(J+1)$ in terms of the Landé factor $g_J=1+\{[J(J+1)+S(S+1)-L(L+1)]/[2J(J+1)]\}$ as function of the spin ($S$), orbital ($L$) and total ($J$) angular momentum quantum numbers. A good fit is achieved using an effective moment value of $\mu_{eff}$ =2.54 $\mu_B$ Ce$^{-1}$. Interestingly, the obtained Weiss temperature is negative; $\theta_P = -30$ K, which suggests dominant *antiferro*magnetic exchange in this compound. However, a steep and *ferro*magnetic-like increase in $\chi(T)$ is found upon cooling below 10 K. The point of steepest incline at this anomaly is marked with an arrow labelled as the Curie point $T_C$=5.2 K in Fig. 6.

The specific heat $C_p$ per formula unit of LT-Ce$_2$Rh$_2$Ga is shown on a logarithmic temperature axis in the main panel of Fig. 7. Near room temperature $C_p \approx$125 J mol$_{fu}^{-1}$ K$^{-1}$ for this compound which closely resembles the phenomenological Dulong-Petit value for a compound with five independent atoms in its formula unit [44]. $C_p$ decreases steadily below room temperature but between 6 K and 5 K $C_p$ rises sharply into a peak of nearly 20 J mol$_{fu}^{-1}$ K$^{-1}$ before decreasing further towards lower temperatures. The calculated configurational entropy $S$ per mole Ce is shown in inset (a) of Fig. 7, where $S(T) = \int_0^T (C_P/T')dT'$. An entropy amount of 4.2 J mol$_{Ce}^{-1}$ K is released at the phase transition which is somewhat lower than the value $R\ln(2)$=5.76 J mol$^{-1}$ K$^{-1}$ attributable to long-range magnetic ordering out of a doublet ground state.

In inset (b) to Fig. 7 specific heat in the form $C_p(T)/T$ is plotted against the square of temperature. The linear behaviour above $T_C$ enables to estimate the electronic contribution on account of the different temperature dependencies of the electronic γ and the lattice β specific heats; $C_p(T)/T = \gamma + \beta T^2$. The solid line on the data in inset (b) is the result of a least-square fit with γ =75 mJ mol$_{Ce}^{-1}$ K$^{-2}$ and β =0.547 mJ mol$_{Ce}^{-1}$ K$^{-4}$.

*HT-Ce$_2$Rh$_2$Ga*. Figure 8 (main panel) shows the per-mole Ce magnetic susceptibility of HT-Ce$_2$Rh$_2$Ga measured in a static field of 0.1 T. The line on the experimental points is the result of fitting the data to the Curie-Weiss law. Between 400 K and 200 K the experimental points may thus be described using an effective magnetic moment value of 2.54 μ$_B$ Ce$^{-1}$ and a Weiss temperature θ$_p$= −120 K. However, upon cooling below the predictable Curie-Weiss range, at a temperature labelled $T_N$ = 128.5 K χ drops precipitously and continues to decrease losing about 25% of its magnitude at $T_N$ when ~80 K is reached. Below this temperature χ rises again towards low temperatures. The shape of the anomaly in χ(T) at $T_N$ is not unlike what may be expected at a paramagnetic-to-antiferromagnetic phase transition, which is corroborated by the dominant antiferromagnetic nature of the magnetic exchange obtained from the calculated negative value of the Weiss temperature for this compound. In the inset of Fig. 8 the region near $T_N$ is shown on expanded scales. Between data collected while slowly cooling the sample (black symbols) and warming the sample (red symbols) a thermal hysteresis amounting to Δ$T_N$ = 1 K is found. Outside of the region of Δ$T_N$ the warming and cooling data curves coincide. Figure 9 shows the results of magnetization measurements at a number of fixed temperatures. The three isotherms in panel (a) were collected at three temperatures close to the phase transition temperature. The field-dependent magnetization in this temperature region proceeds linearly in field up to 7 T. The magnetic moment extracted from magnetization measurements even at low temperature and deep into the ordered region is small – see Fig. 9(b). An applied field of 9 T yields only 0.34 μ$_B$ Ce$^{-1}$ for the sample at 1.7 K. It is noteworthy, however, that correlated electron Ce-based compounds often yield only small magnetization values [45, 46].

Finally, we turn to the specific heat of HT-$Ce_2Rh_2Ga$ as illustrated in the main panel of Fig. 10 on a per-chemical formula unit basis. Below room temperature we notice that $C_p$ is nearly constant at a value close to the Dulong-Petit value for this compound. In the Debye model of the specific heat of solids this value is achieved when the substance is at a temperature suitably higher than its Debye temperature $\theta_D$. At 130 K however, the monotonous behavior in $C_p(T)$ ceases where $C_p$ rises very sharply and achieves 424 J $mol_{fu}^{-1}$ $K^{-1}$ at 127.5 K. In the inset of Fig. 10 we cast the specific heat data below 20 K in the form $C_p(T)/(T)$ against the square of temperature. The line on the experimental points is drawn with $\gamma$=172.5 mJ $mol_{Ce}^{-1}$ $K^{-1}$ and $\beta$=1.217 x $10^{-3}$ mJ $mol_{Ce}^{-1}$ $K^{-4}$. By contrast, in the specific heat of the comparable nonmagnetic compound $La_2Rh_2Ga$ (see Fig. 11a) there are no anomalies observed below room temperature, as would be expected for a simple metal. For $La_2Rh_2Ga$ the Sommerfeld coefficient is found to be $\gamma$ =13.45 mJ $mol_{La}^{-1}$ $K^{-2}$ according to the least-squares fit shown in panel (b) of Fig. 11.

The unusual behavior of HT-$Ce_2Rh_2Ga$ at 128.5 K prompted us to search for a possible crystal structure phase transition. The first attempts to determine structural changes in the transition through $T_N$=128 K using single crystal X-ray diffraction showed that the structure changes are too small to be reliably described quantitatively. Changes in cell dimensions and cell volume as well as the appearance of a small monoclinic distortion when cooling a single crystal of HT-$Ce_2Rh_2Ga$ from room temperature down to 80 K are presented in Figure S3 (Supplementary materials). Further crystal structure determination was not possible due to the twinning of the crystals induced by the structural transition.

**4 Discussion**

The two new Ce-based compounds LT-$Ce_2Rh_2Ga$ and HT-$Ce_2Rh_2Ga$ show contrasting behaviours despite being alike in their chemical compound formulas. In both structures Ce atoms occupy a unique crystallographic site however in LT-$Ce_2Rh_2Ga$ it is a general position 8*f* (site symmetry 1) and in HT-$Ce_2Rh_2Ga$ it is a special position 8*f* (site symmetry *m*). Magnetically, both compounds can be described in terms of the full trivalent magnetic state of the Ce ion over a wide range below room temperature, which makes both compounds amenable to magnetic order. In LT-$Ce_2Rh_2Ga$ the Ce ions order in a spin arrangement at a temperature of 5.2 K that imparts a ferromagnetic character to its temperature-dependent magnetic susceptibility as well as its field-dependent magnetization. This magnetic phase transition temperature is typically in the range of many cerium-based compounds. The single *4f*-electron of cerium in its trivalent state being predisposed to an antiferromagnetic on-site Kondo exchange in metals, however, means that by comparison ferromagnetic order in cerium intermetallic compounds is much less ubiquitous than antiferromagnetic order.

Taken together, the results of magnetic susceptibility and specific heat for the phase transition found in this study on the compound HT-Ce$_2$Rh$_2$Ga bear the signatures of antiferromagnetic order, but at the extraordinary high phase transition temperature of 128.5 K. With the present state of investigation and results available on this compound we cannot rule out the possibility of a crystal structure modification accompanying the magnetic phase transition at $T_N$. This phase transition is temperature hysteretic furthermore which is unusual since fluctuations in the order parameter is a common feature of magnetic order driven by an anti-parallel spin arrangement.

In conclusion, the novel Ce$_2$Rh$_2$Ga intermetallics crystallize in two polymorphs adopting rather unusual structure-types: an orthorhombic ordered La$_3$Ni$_2$-type for samples annealed above 867°C and a monoclinic Pr$_2$Co$_2$Al-type for samples annealed below this temperature. Both structure types can be built from CeGa$_4$ cerium centered gallium tetrahedral bricks. Both allotropes show phase transitions of a magnetic nature and further studies are needed to determine the order parameter. In HT-Ce$_2$Rh$_2$Ga in particular it would be interesting to clarify what property of the material is driving the temperature hysteresis that accompanies the magnetic ordering.

**Acknowledgements**

AMS thanks the URC/FRC of UJ and the SA NRF (93549) for generous financial assistance, and the Max Planck Institute CPfS in Dresden for their hospitality where parts of this research were conducted. This study was supported by the Russian Foundation for Basic Research (Grant No 18-03-00656a and 19-03-00135).

Table 1. Selected single-crystal data collection and structure refinement parameters for La$_2$Rh$_2$Ga , HT-Ce$_2$Rh$_2$Ga, and LT-Ce$_2$Rh$_2$Ga.

| Empirical formula | La$_2$Rh$_2$Ga | HT-Ce$_2$Rh$_2$Ga | LT-Ce$_2$Rh$_2$Ga |
|---|---|---|---|
| Molar mass, g·mol$^{-1}$ | 553.36 | 555.78 | |
| Space group | *Cmce* (64) | *Cmce* (64) | *C* 2/*c* (15) |
| Structure type | La$_2$Ni$_3$ | La$_2$Ni$_3$ | Pr$_2$Co$_2$Al |
| *a* (Å) | 5.9251(17) | 5.851(2) | 10.0903(6) |
| *b* (Å) | 9.8402(17) | 9.618 (2) | 5.6041(3)  β=104.995(3)° |
| *c* (Å) | 7.5019(14) | 7.487 (2) | 7.8153(4) |
| Cell volume (Å$^3$) | 437.39(17) | 421.3(2) | 426.88(4) |
| Z | 4 | 4 | 4 |
| D calc (g*cm$^{-3}$) | 8.403 | 8.762 | 8.648 |
| Abs. coeff μ (mm$^{-1}$) | 17.046 | 18.437 | 34.464 |
| Radiation (Å) | AgKα, 0.56087 | AgKα, 0.56087 | MoKα, 0.71073 |
| Index range | -8≤ *h* ≤8, 0≤ *k* ≤14, 0≤ *l* ≤11 | 0≤ *h* ≤11, -19≤ *k* ≤0, -15≤ *l* ≤15 | -16≤ *h* ≤16, 0≤ *k* ≤9, 0≤ *l* ≤13 |
| *Θ* range | 3.27 – 24.5 | 3.34° - 34.96° | 4.18° - 36.53° |
| Number of measured reflections | 803 | 983 | 1060 |
| Number of reflections with I ≥ 2σ(I) | 361 | 597 | 937 |
| Number of refined parameters | 17 | 17 | 25 |
| GooF on $F^2$ | 1.116 | 0.972 | 1.088 |
| $R[F^2>2\sigma(F^2)]$ | 0.0310 | 0.0389 | 0.0239 |
| $wR(F^2)$ | 0.0843 | 0.0641 | 0.0537 |

Table 2. Atomic coordinates and equivalent isotropic displacement parameters $U_{eq}$ (Å$^2$) for La$_2$Rh$_2$Ga , HT-Ce$_2$Rh$_2$Ga, and LT-Ce$_2$Rh$_2$Ga.

| Atom | Wyckoff site | x/a | y/b | z/c | U$_{eqv}$ |
|---|---|---|---|---|---|
| La$_2$Rh$_2$Ga | | | | | |
| La | 8f | 0 | 0.33857(4) | 0.09757(6) | 0.0132(2) |
| Rh | 8e | 1/4 | 0.09407(5) | 1/4 | 0.0146(2) |
| Ga | 4a | 0 | 0 | 0 | 0.0128(3) |
| HT-Ce$_2$Rh$_2$Ga | | | | | |
| Ce | 8f | 0 | 0.33904(4) | 0.09881(6) | 0.0123(1) |
| Rh | 8e | 1/4 | 0.09769(6) | 1/4 | 0.0143(2) |
| Ga | 4a | 0 | 0 | 0 | 0.0118(2) |
| LT-Ce$_2$Rh$_2$Ga | | | | | |
| Ce | 8f | 0.35324(2) | 0.14759(4) | 0.35191(3) | 0.01041(7) |
| Rh | 8f | 0.13144(3) | 0.13464(5) | 0.00765(4) | 0.01100(8) |
| Ga | 4e | 0 | 0.1398(2) | 1/4 | 0.0098(2) |

Table 3. Interatomic distances (Å) in the HT- and LT- Ce$_2$Rh$_2$Ga

| HT- Ce$_2$Rh$_2$Ga | | | LT- Ce$_2$Rh$_2$Ga | | | | | |
|---|---|---|---|---|---|---|---|---|
| Atom | To atom | $d$ | Atom | To atom | $d$ | Atom | To atom | $d$ |
| Ce | 2 Rh | 2.9680(8) | Ce | Rh | 2.9310(4) | Rh | Ga | 2.5781(3) |
|  | 2 Rh | 3.0545(7) |  | Rh | 3.0227(4) |  | Ga | 2.6070(5) |
|  | 2 Rh | 3.1000(8) |  | Rh | 3.0227(4) |  | Rh | 2.7494(6) |
|  | Ga | 3.3438(7) |  | Rh | 3.0666(4) |  | Ce | 2.9310(4) |
|  | Ga | 3.3793(8) |  | Rh | 3.1040(4) |  | Ce | 3.0137(4) |
|  | 2 Ga | 3.391(1) |  | Rh | 3.2252(4) |  | Ce | 3.0227(4) |
|  | Ce | 3.432(1) |  | Ga | 3.3028(3) |  | Rh | 3.0270(6) |
|  | 2 Ce | 3.699(1) |  | Ga | 3.3235(6) |  | Ce | 3.0667(4) |
|  | 2 Ce | 3.699(2) |  | Ga | 3.3965(6) |  | Ce | 3.1040(4) |
| Rh | 2 Ga | 2.5546(5) |  | Ga | 3.4455(3) |  | Ce | 3.2252(4) |
|  | 2 Rh | 2.925 (2) |  | Ce | 3.6072(3) | Ga | 2 Rh | 2.5782(4) |
|  | 2 Ce | 2.9680(8) |  | Ce | 3.6073(3) |  | 2 Rh | 2.6071(5) |
|  | 2 Ce | 3.0545(7) |  | Ce | 3.6495(5) |  | 2 Ce | 3.3028(3) |
|  | 2 Ce | 3.1000(8) |  | Ce | 3.6759(5) |  | 2 Ce | 3.3235(6) |
| Ga | 4 Rh | 2.5546(5) |  | Ce | 3.7081(5) |  | 2 Ce | 3.3965(6) |
|  | 2 Ce | 3.3438(8) |  |  |  |  | 2 Ce | 3.4455(3) |
|  | 2 Ce | 3.3793(8) |  |  |  |  |  |  |
|  | 4 Ce | 3.391 (1) |  |  |  |  |  |  |

**Figure Captions**

Fig. 1 DTA curve of the annealed at 700°C sample (the arrows show the course of temperature change).

Fig. 2 Rietveld refined X-ray diffraction patterns for (up) HT-$Ce_2Rh_2Ga$ ($\chi^2$ = 1.78, $R_p$ = 0.024, $R_{exp}$ = 0.022) and (down) LT-$Ce_2Rh_2Ga$ ($\chi^2$ = 3.69, $R_p$ = 0.024, $R_{exp}$ = 0.015). The experimental diffraction profile is indicated by black dots. The calculated diffraction profile is shown as the upper blue line, the difference profile is shown as the bottom red line and the vertical bars correspond to the calculated Bragg positions.

Fig.3. Coordination polyhedra of the three atoms in the crystal structure of HT-$Ce_2Rh_2Ga$ (upper row) and LT-$Ce_2Rh_2Ga$ (bottom row). Ce-atoms are drawn as large green balls, Rh-atoms as small blue ball, and Ga-atoms as middle size pink balls.

Fig. 4. (a) $CeGa_4$ and (b) $RhGa_4$ tetrahedra arrangement in the unit cell of HT-$Ce_2Rh_2Ga$. (c) The packing of the $CeGa_4$ and $RhGa_4$ tetrahedra. $CeGa_4$ tetrahedra are highlighted in green, $RhGa_4$ tetrahedra are highlighted in blue.

Fig. 5. (a) $CeGa_4$ tetrahedra and (b) double centered Rh octahedra arrangement in the unit cell of LT-$Ce_2Rh_2Ga$. (c) The packing of the $CeGa_4$ tetrahedra and $Rh_2Ga_6$ octahedra. $CeGa_4$ tetrahedra are highlighted in green. $Rh_2Ga_6$-octahedra are emphasized in blue.

Fig. 6. (main panel) Semi-log plot of dc-magnetic susceptibility per mole Ce of LT-$Ce_2Rh_2Ga$ measured in a static applied field of 0.1 T, with the arrow marking the paramagnetic-to-ferromagnetic phase transition temperature $T_C$. (inset) Inverse susceptibility against temperature. The solid line is a Curie-Weiss fit to the data as explained in the text.

Fig. 7. (main panel) Semi-log plot of the specific heat $C_p$ per mole of Ce of LT-$Ce_2Rh_2Ga$. The arrow indicates the ferromagnetic phase transition temperature. Dashed line is a guide to the eye connecting the experimental points. Inset (a): Calculated entropy per mole of Ce. Inset (b): Specific heat in the form $C_p(T)/T$ against the square of temperature. The dashed line is a guide to the eye. Solid line illustrates a fit to extract the Sommerfeld coefficient $\gamma$ = 75 mJ mol$_{Ce}^{-1}$ K$^{-2}$.

Fig. 8. (main panel) Magnetic dc-susceptibility of HT-$Ce_2Rh_2Ga$ measured in a static applied field of 0.1 T. The solid line on the data is obtained through a Curie-Weiss fit. The dashed line illustrates the continuation of the Curie-Weiss fit below the fitted temperature range which is intercepted at the phase transition marked $T_N$ at 128.5 K. (inset) Magnetic susceptibility on expanded scales near $T_N$ shows the temperature hysteretic behaviour between data obtained during cooling (black symbols) and warming (red symbols) of the sample.

Fig. 9. dc-Magnetization per Ce atom of HT-$Ce_2Rh_2Ga$ at temperatures (a) close to $T_N$ and (b) at low temperature. The magnetization is linear in field at all investigated temperatures except for the $T$=1.70 K isotherm where the high-field region permits to calculate a saturation magnetization of 0.19 $\mu_B$/Ce.

Fig. 10. (main panel) Temperature dependence of specific heat per formula unit of HT-$Ce_2Rh_2Ga$. The dashed line is a guide to the eye to connect the data points through the sharp phase transition marked by an arrow at $T_N$ =128.5 K. (inset) Specific heat per mole of Ce of HT-$Ce_2Rh_2Ga$ in the form $C_p(T)/T$ plotted against the square of temperature. The solid line on the data is a fit to extract the electronic specific heat coefficient $\gamma$=172.5 mJ $mol_{Ce}^{-1}K^{-2}$. An upturn of unknown origin is found to occur in $C_p(T)/T$ below about 7 K.

Fig. 11. (a) Specific heat of the Pauli-paramagnetic compound $La_2Rh_2Ga$ over a wide range of temperature. (b) Specific heat in the form $C_p(T)/T$ against the square of temperature. A value $\gamma$=13.45 mJ $mol_{La}^{-1}$ $K^{-2}$ is obtained for the Sommerfeld coefficient as illustrated by the solid line on the data points.

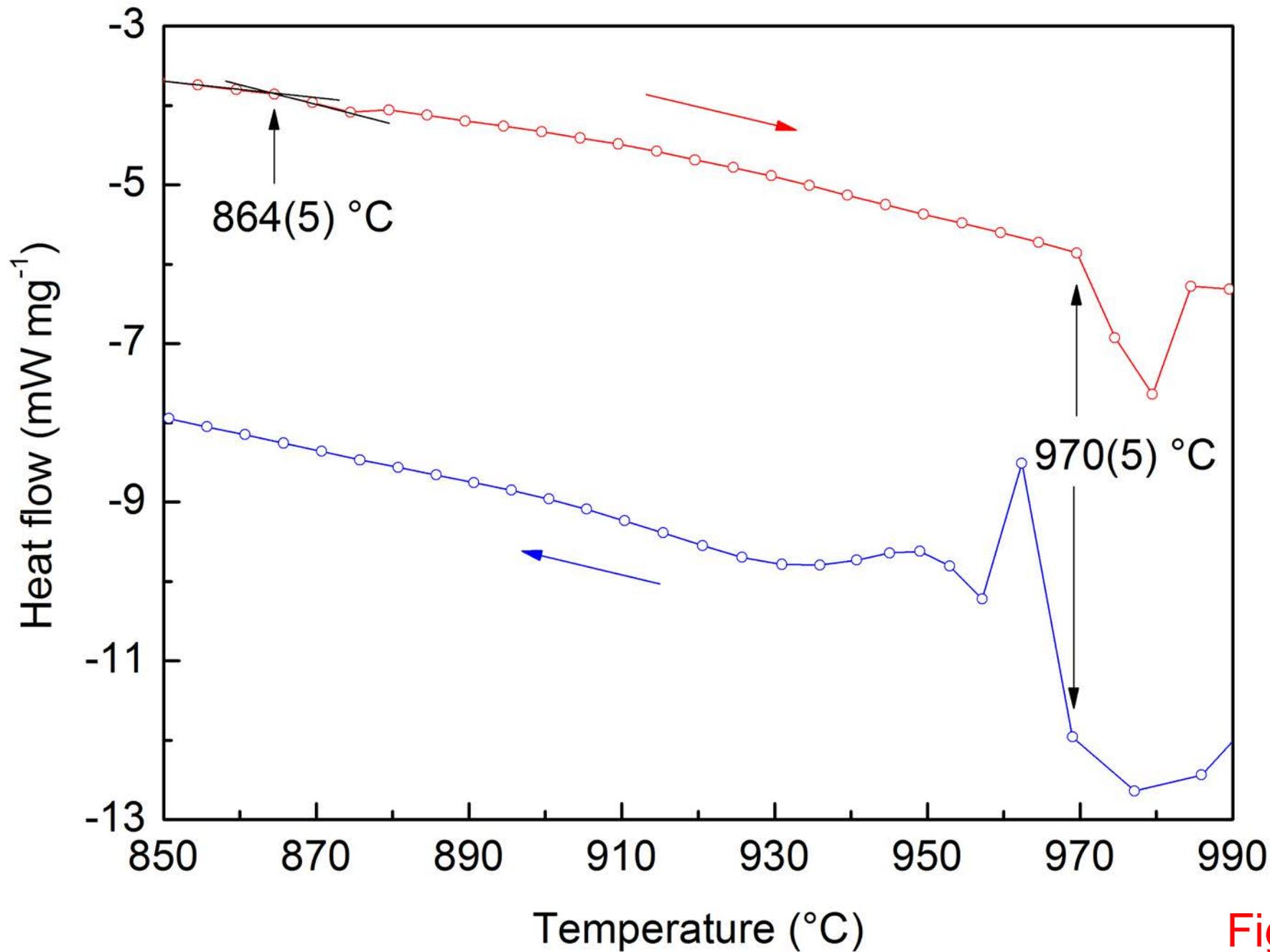

Figure 1

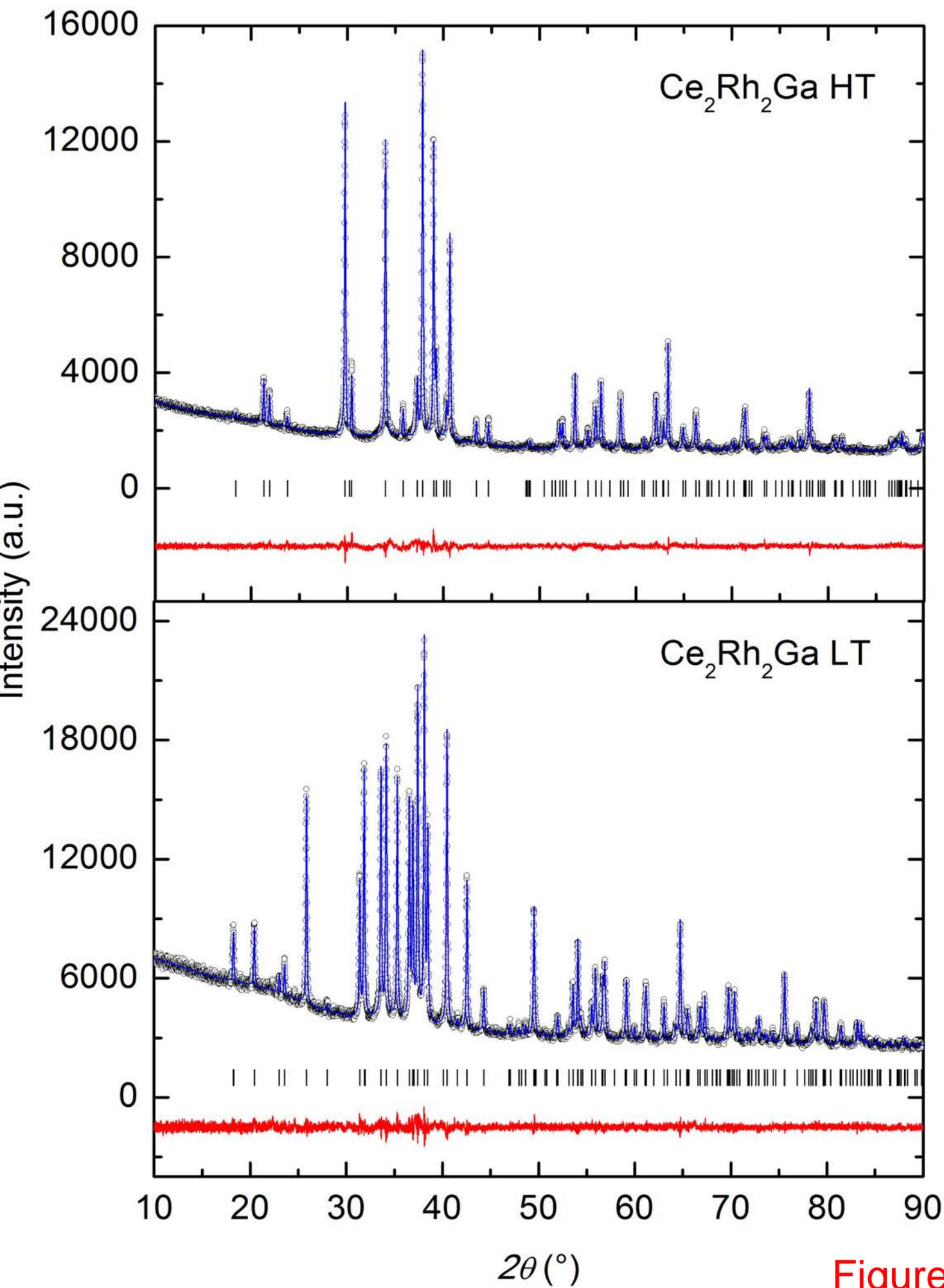

Figure 2

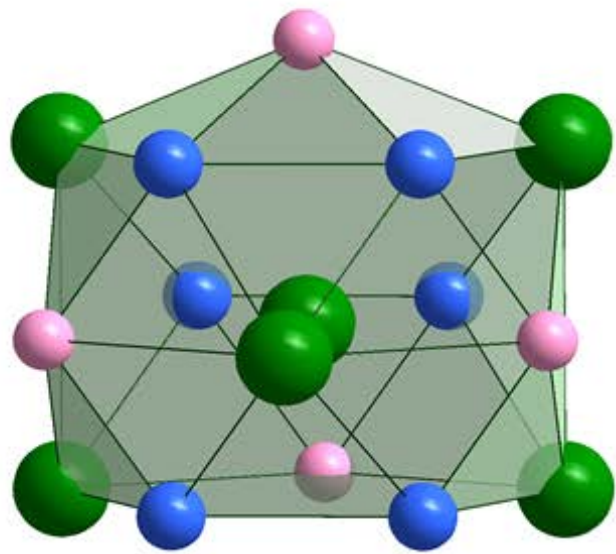
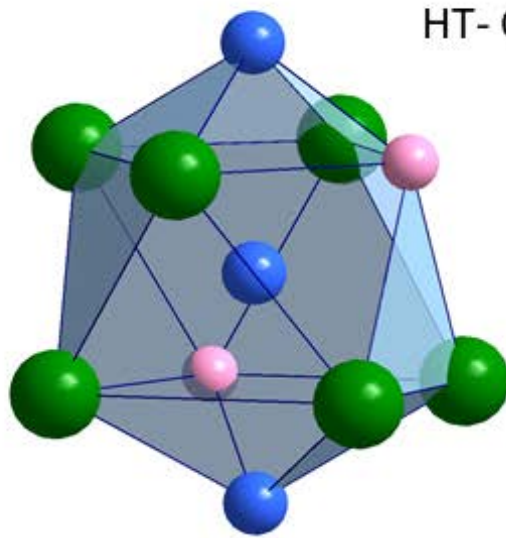
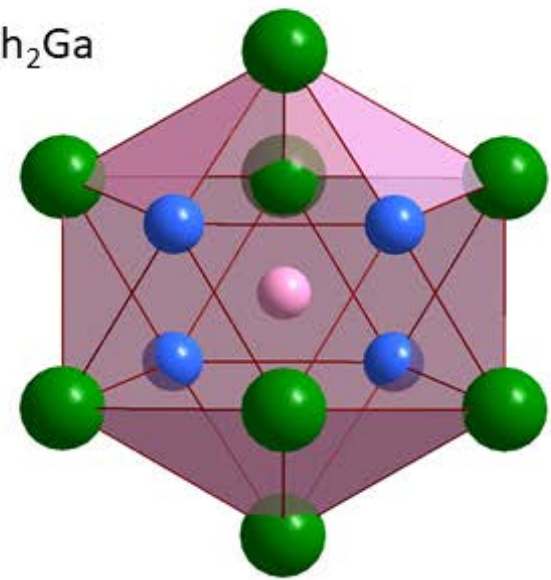
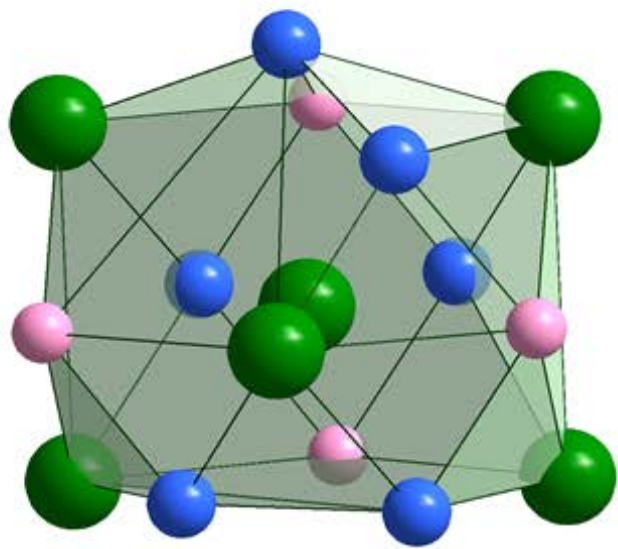
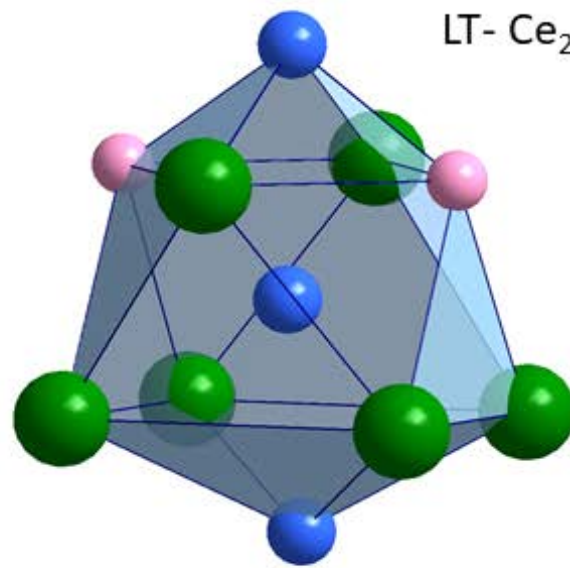
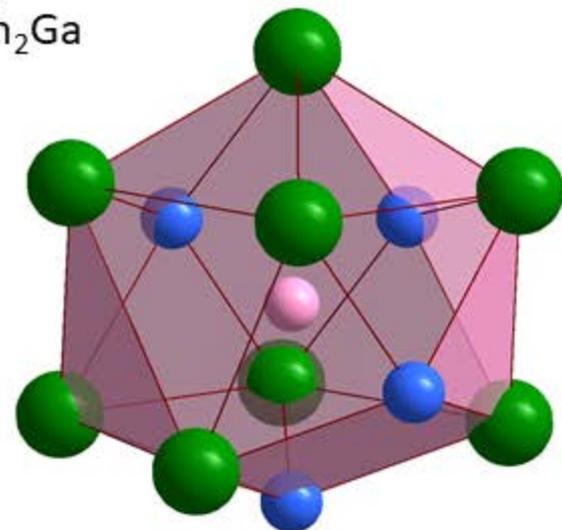

Figure 3

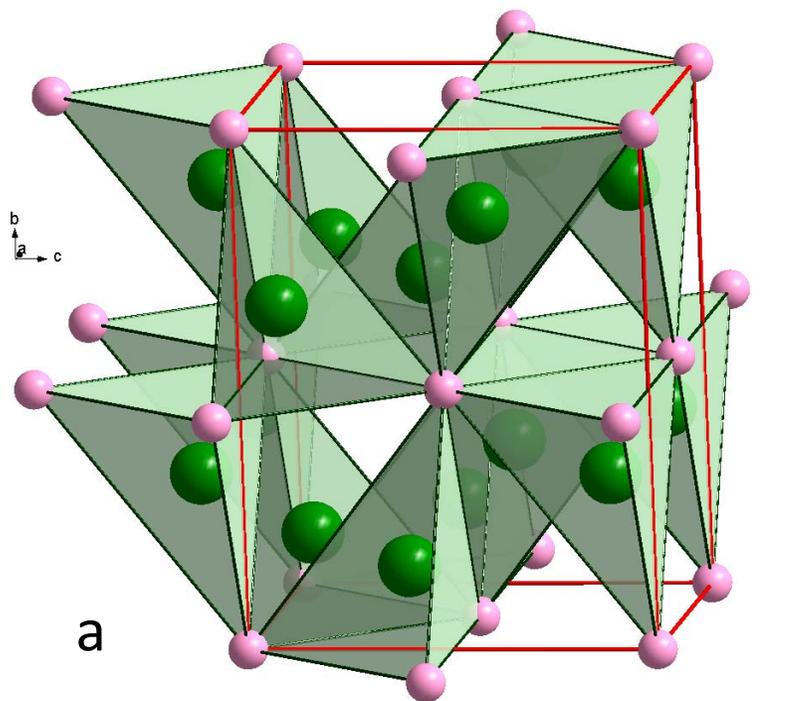
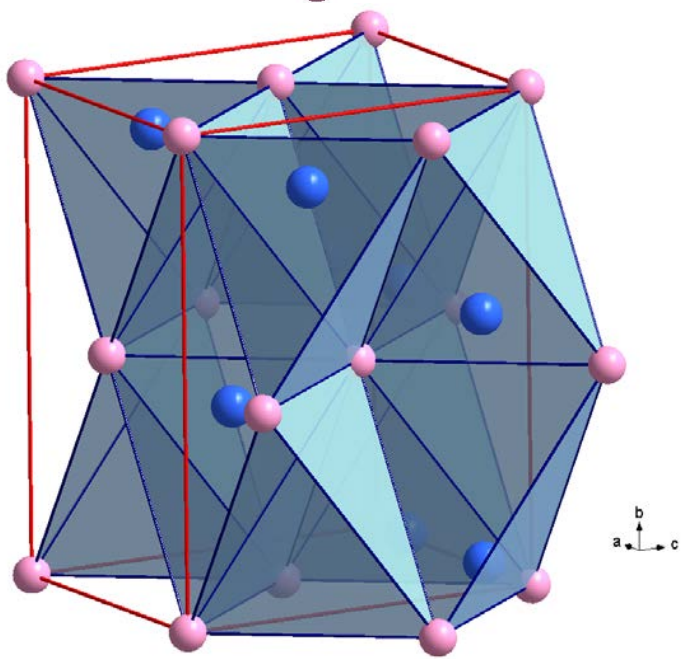
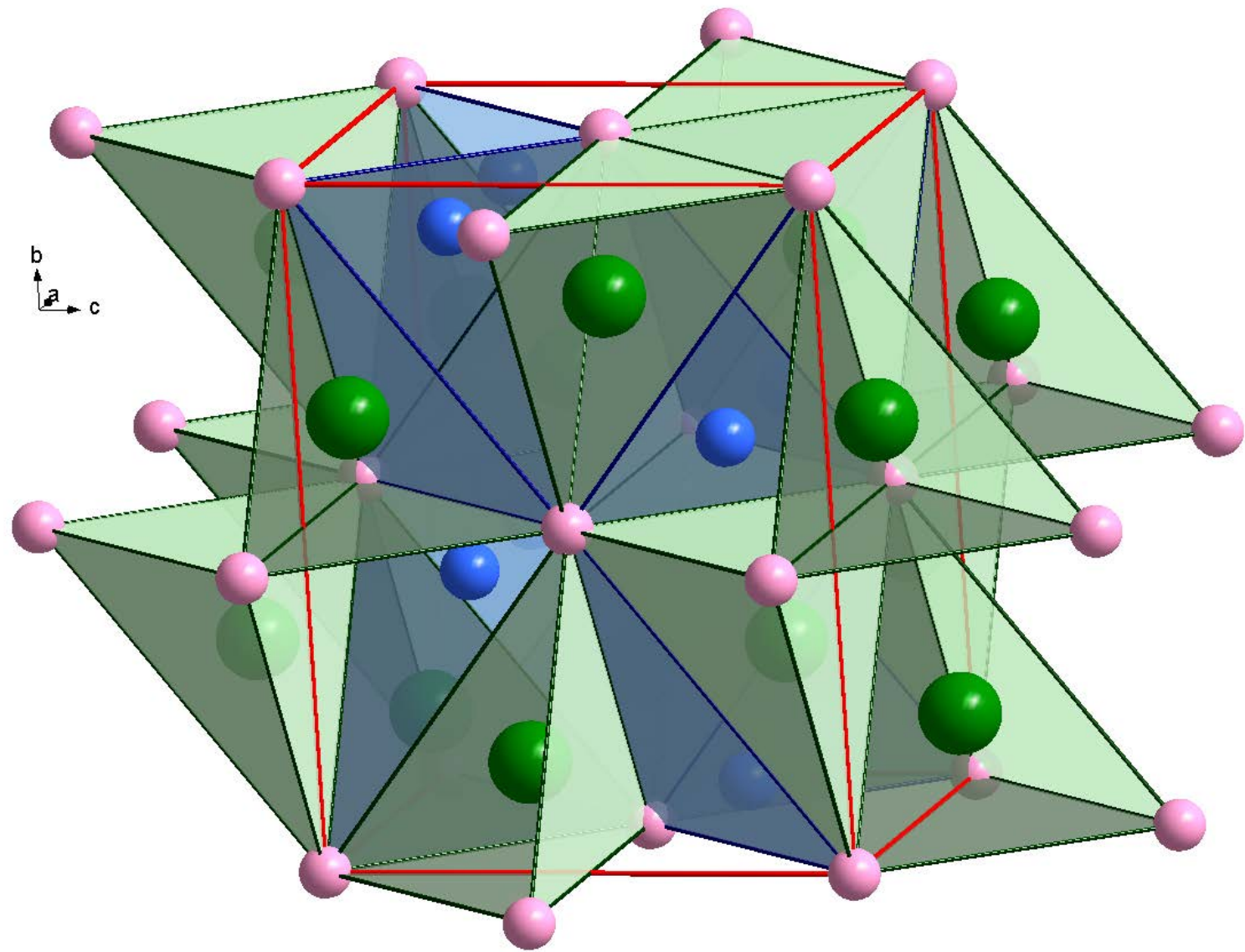

Figure 4

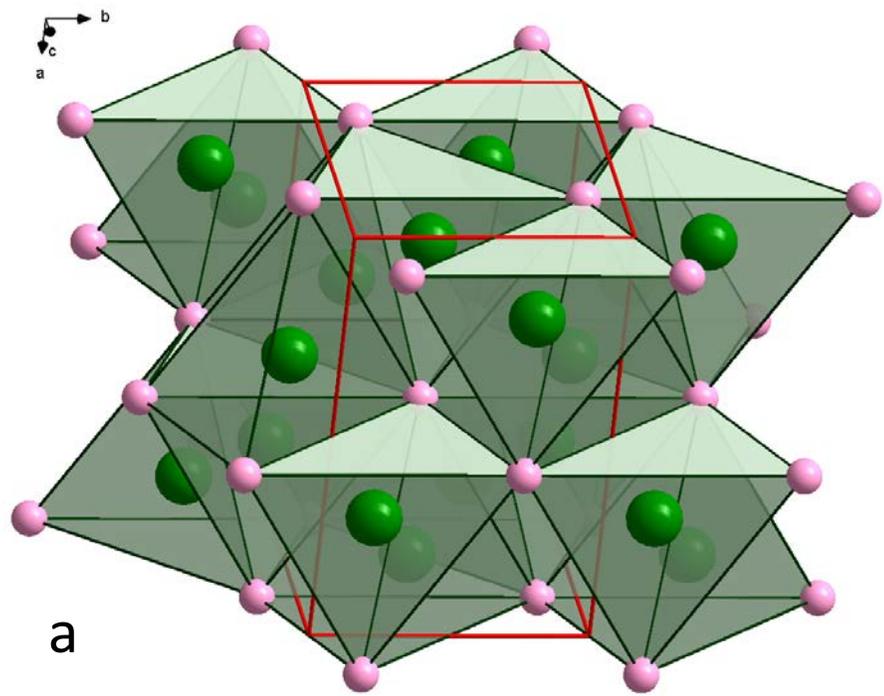
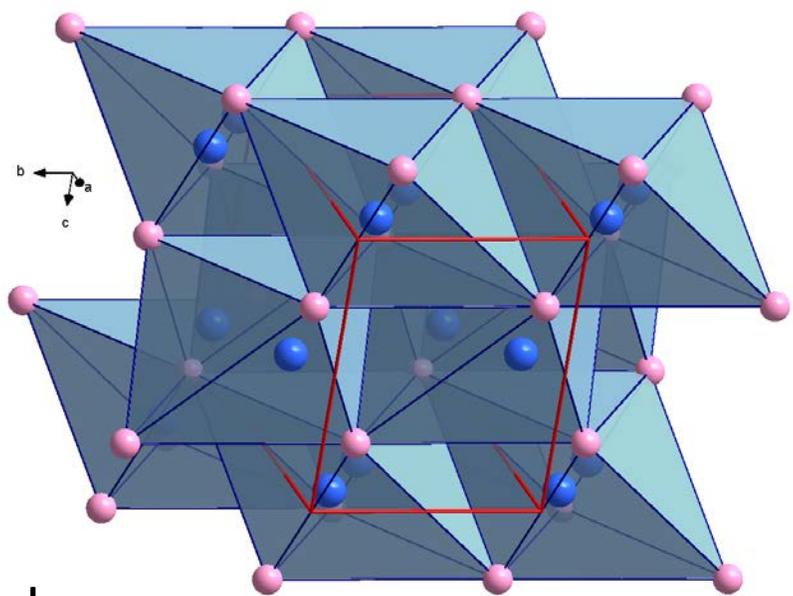
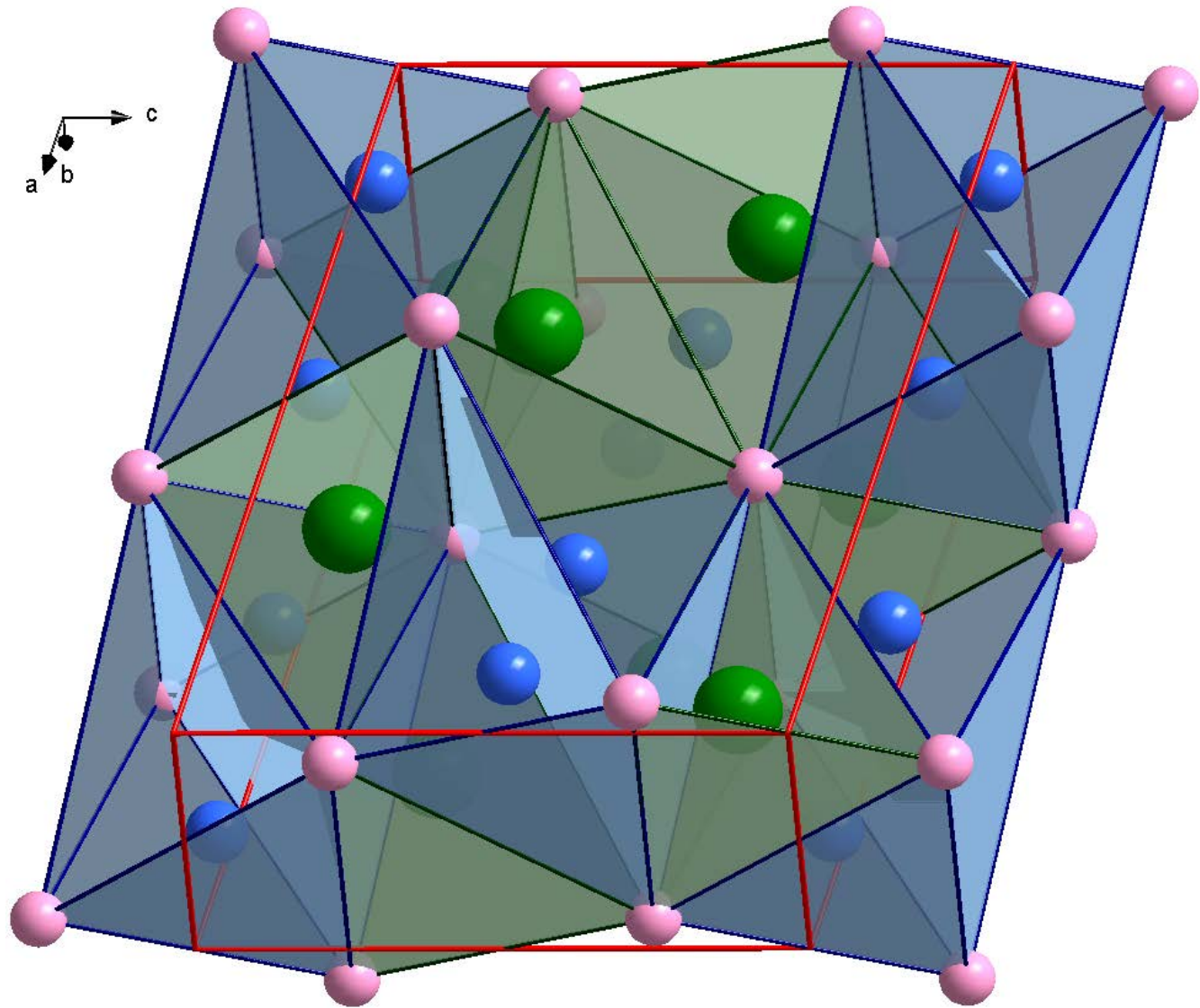

Figure 5

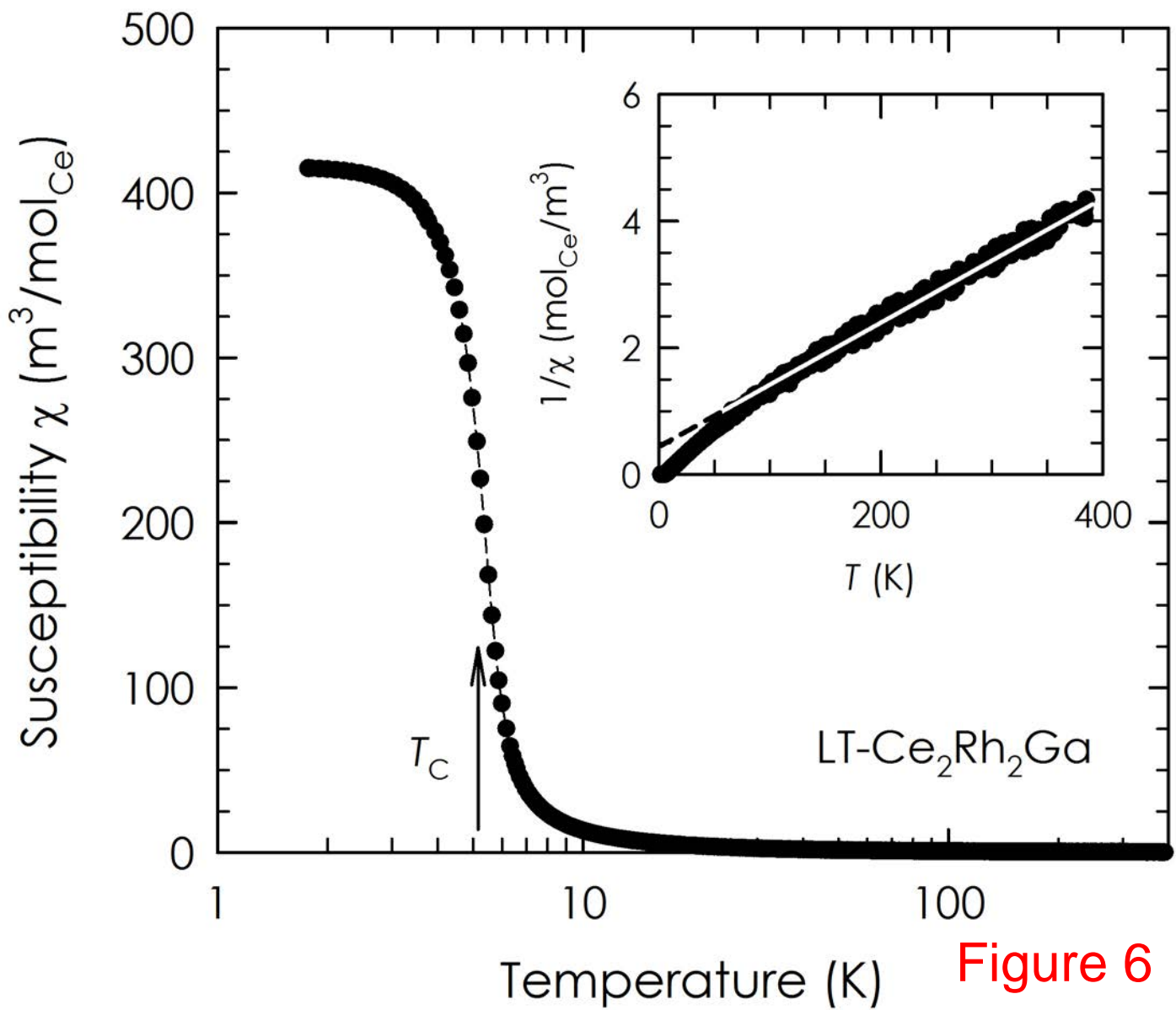

Figure 6

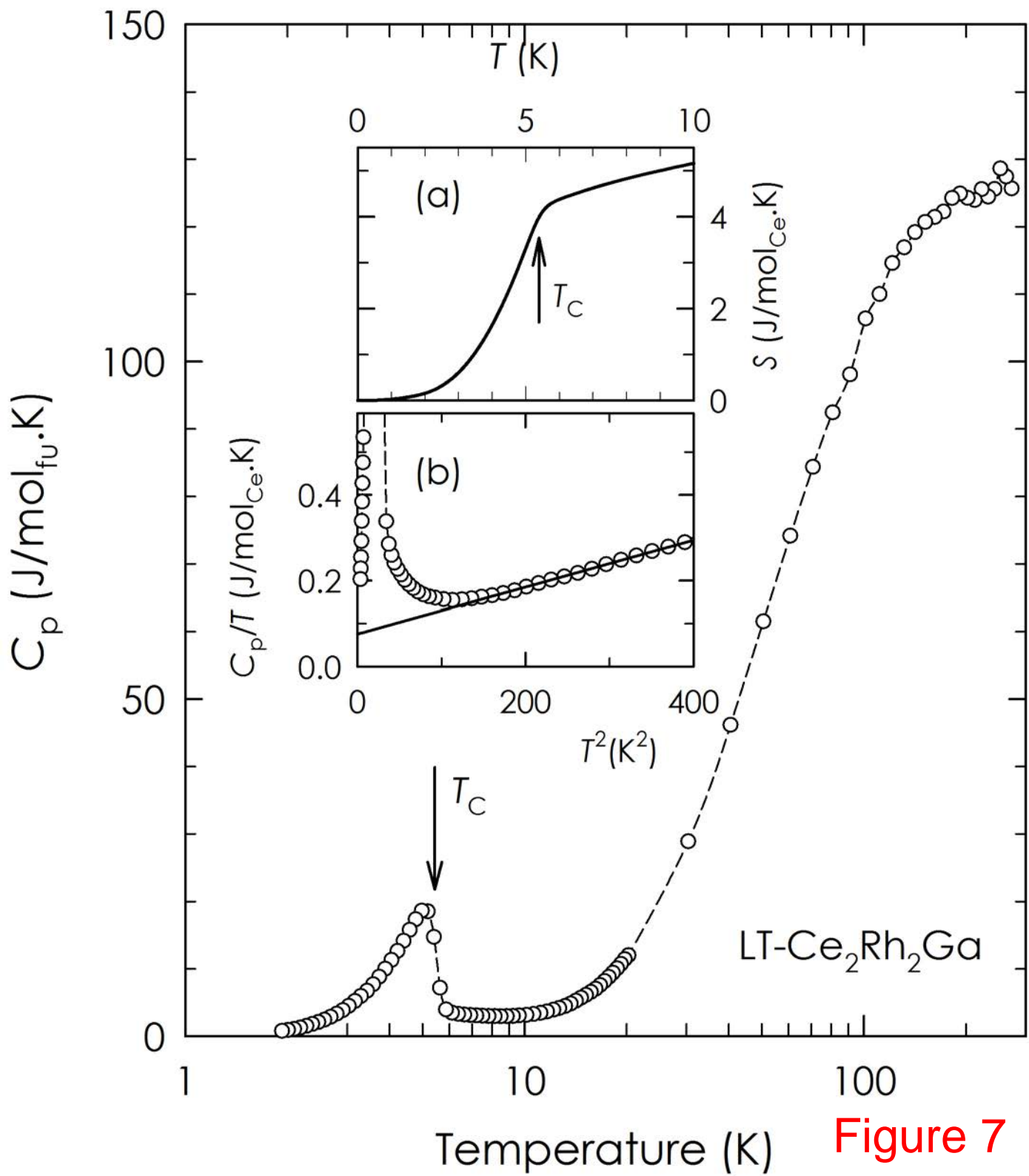

Figure 7

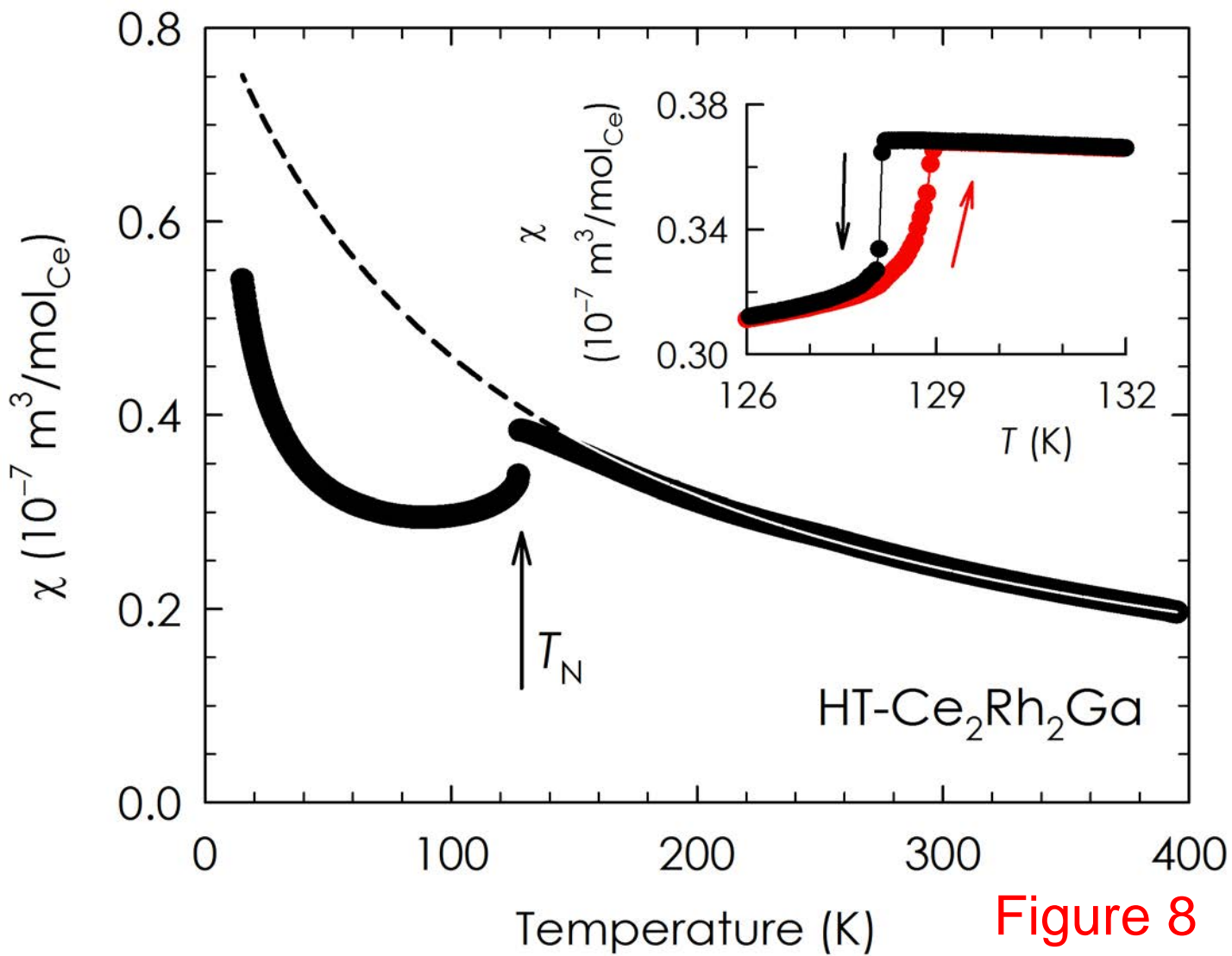

Figure 8

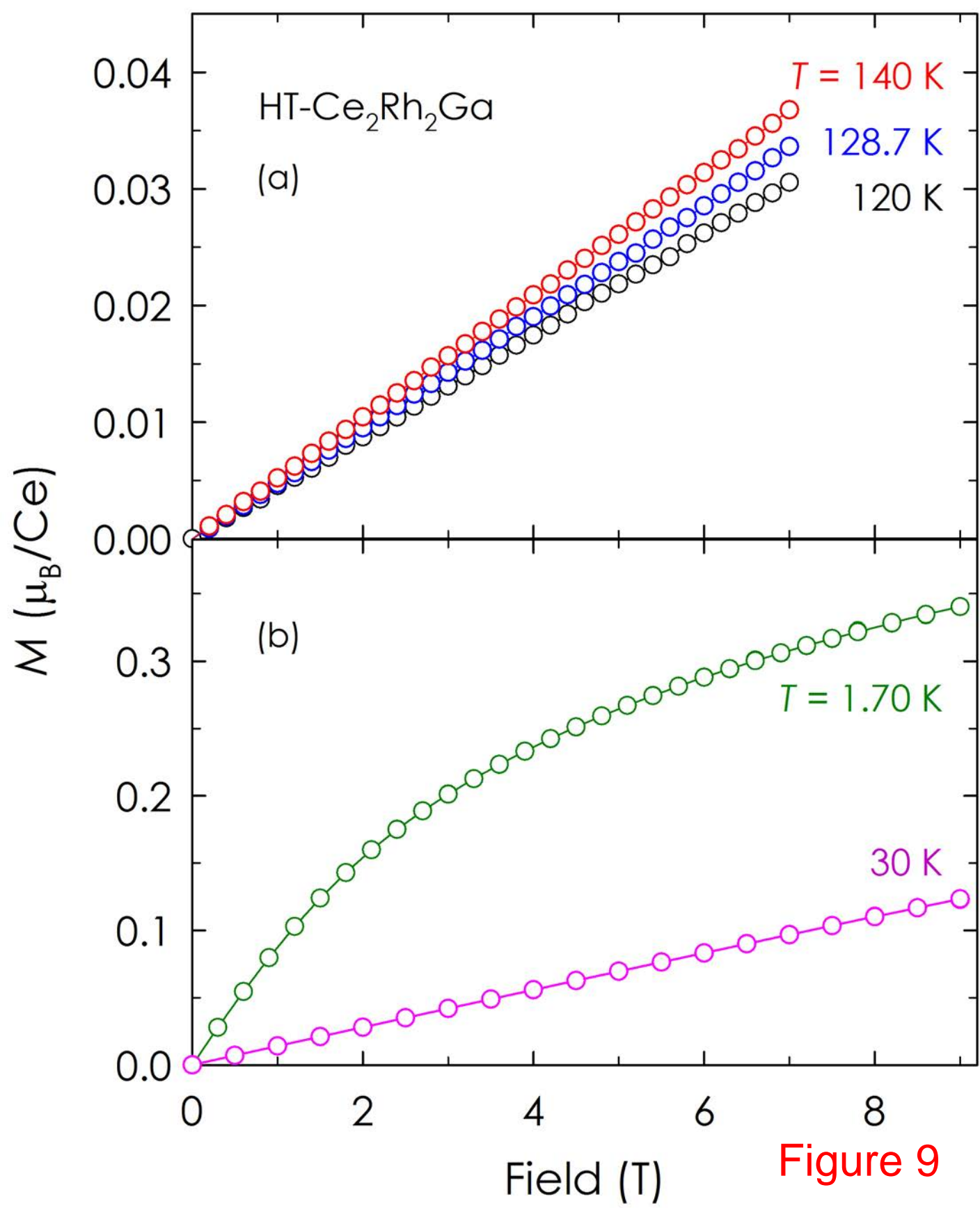

Figure 9

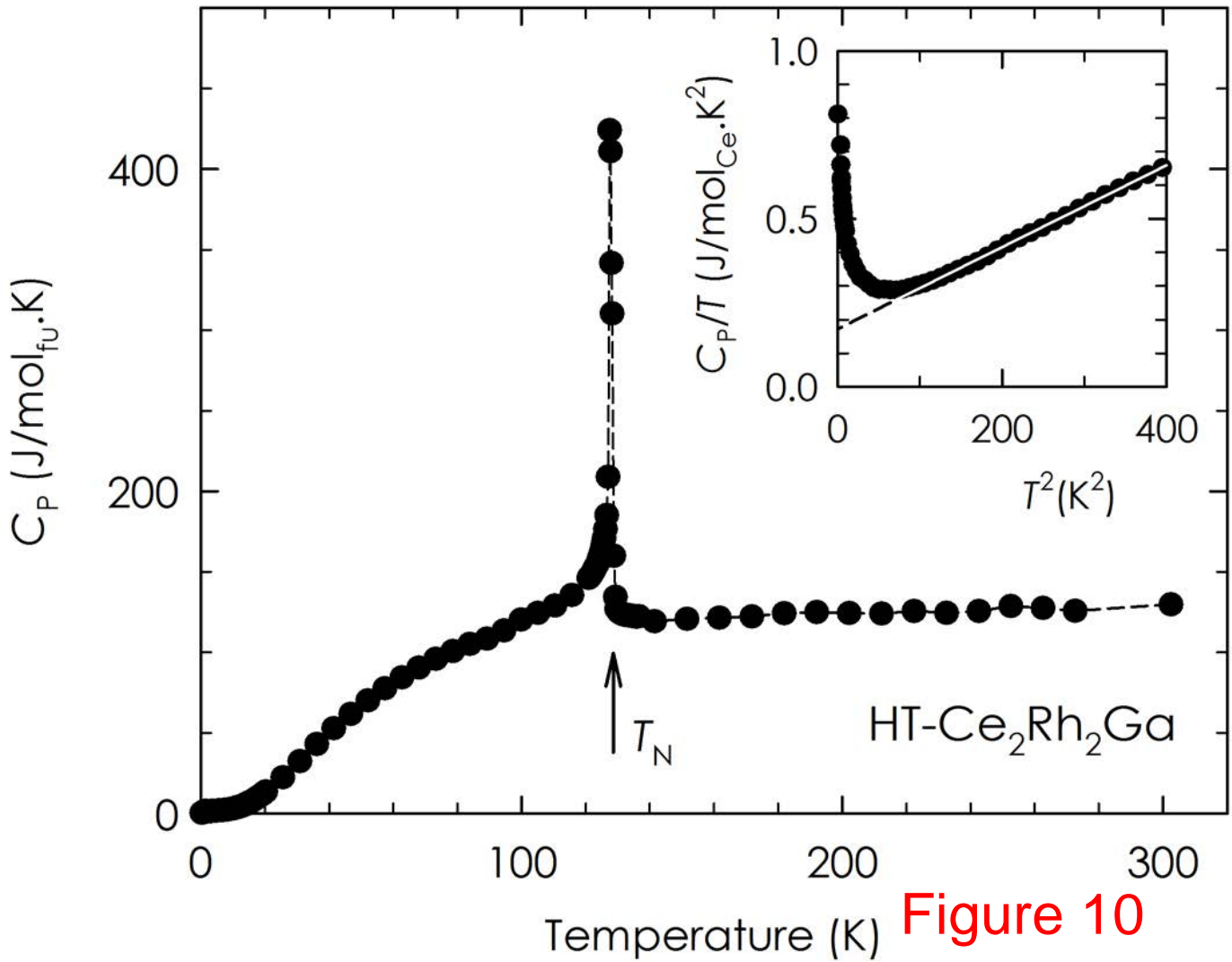

Figure 10

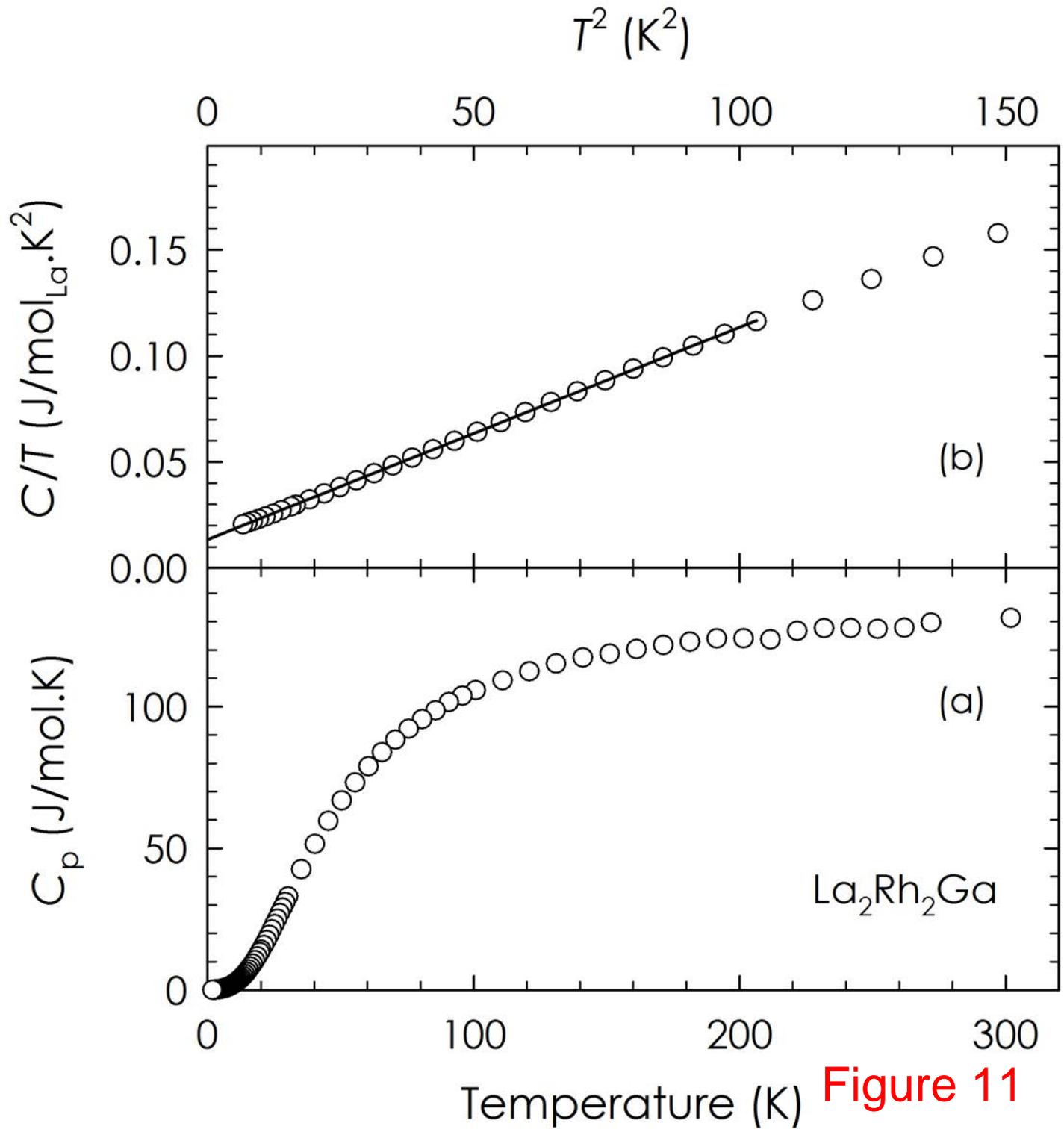

Figure 11